\documentclass[conference]{IEEEtran}
\IEEEoverridecommandlockouts
\usepackage{cite}
\usepackage{amsmath,amssymb,amsfonts}
\usepackage{algorithmic}
\usepackage{graphicx}
\usepackage{textcomp}
\usepackage{xcolor}
\usepackage{multirow}
\usepackage{graphicx, subfigure, epsfig}
\def\BibTeX{{\rm B\kern-.05em{\sc i\kern-.025em b}\kern-.08em
    T\kern-.1667em\lower.7ex\hbox{E}\kern-.125emX}}
\begin{document}

\title{A Matrix Factorization Based Network Embedding Method for DNS Analysis}

\author{\IEEEauthorblockN{Meng Qin}
\IEEEauthorblockA{\textit{Independent Researcher}}
}

\maketitle

\begin{abstract}
In this paper, I explore the potential of network embedding (a.k.a. graph representation learning) to characterize DNS entities in passive network traffic logs. I propose an MF-DNS-E (\underline{M}atrix-\underline{F}actorization-based \underline{DNS} \underline{E}mbedding) method to represent DNS entities (e.g., domain names and IP addresses), where a random-walk-based matrix factorization objective is applied to learn the corresponding low-dimensional embeddings.
\end{abstract}


\section{Methodology}\label{sec:meth}

I introduce an MF-DNS-E (\underline{M}atrix-\underline{F}actorization-based \underline{DNS} \underline{E}mbedding) method to automatically learn the low-dimensional vector representations (i.e., embeddings) for domain names and IP addresses, which can be used to support some DNS security analysis tasks including malicious domain detection and IP reputation evaluation.

Le $N$ and $M$ be the total numbers of (\romannumeral1) domain names and (\romannumeral2) IP addresses in a passive network traffic log, respectively. One can extract a bipartite graph ${\bf{B}} \in {\mathbb{Z}_+ ^{N \times M}}$, with  ${{\bf{B}}_{ij}}$ as the number of queries that a domain name ${p_i}$ is resolved to an IP address ${q_j}$. The corresponding adjacency matrix of this bipartite graph is
\begin{equation}\label{eq:pass}
    {\bf{P}} := \left[ {\begin{array}{*{20}{c}} {\bf{0}}_{N \times M} &{{\bf{\hat B}}}\\ {{\bf{\hat B}}^T} & {\bf{0}}_{M \times N} \end{array}} \right],
\end{equation}
where each entry in ${\bf{B}}$ is normalized into $[0, 1]$ via ${{{\bf{\hat B}}}_{ij}} = {{{{\bf{B}}_{ij}}} \mathord{\left/ {\vphantom {{{{\bf{B}}_{ij}}} {{{\bf{B}}_{\max }}}}} \right. \kern-\nulldelimiterspace} {{{\bf{B}}_{\max }}}}$, with ${\bf{B}}_{\max}$ as the maximum entry.

For all the domain names $\{ {p_i} \}$, I further introduce (\romannumeral1) host-based and (\romannumeral2) IP-based similarities between each pair of domain names, which are denoted as ${{\bf{S}}_{{\rm{DH}}}} \in {\mathbb{R} ^{N \times N}}$ and ${{\bf{S}}_{{\rm{DI}}}} \in {\mathbb{R} ^{M \times M}}$, respectively. Given a domain name ${p_i}$, let ${H_{p_{i}}}$ be the set of hosts that have the DNS query result of ${p_i}$. I further denote ${I_{p_{i}}}$ as the set of IPs resolved from ${p_i}$. ${{\bf{S}}_{{\rm{DH}}}}$ and ${{\bf{S}}_{{\rm{DI}}}}$ are defined as:
\begin{equation}
    {({{\bf{S}}_{{\rm{DH}}}})_{ij}} := \frac{{|{H_{{p_i}}} \cap {H_{{p_j}}}|}}{{|{H_{{p_i}}} \cup {H_{{p_j}}}|}},{({{\bf{S}}_{{\rm{DI}}}})_{ij}} :=  \frac{{|{I_{{p_i}}} \cap {I_{{p_j}}}|}}{{|{I_{{p_i}}} \cup {I_{{p_j}}}|}}.
\end{equation}
For all the IP addresses $\{ {q_j} \}$, I also define the (\romannumeral3) host-based and (\romannumeral4) domain-based similarities, denoted as ${{\bf{S}}_{{\rm{IH}}}} \in {\mathbb{R} ^{M \times M}}$ and ${{\bf{S}}_{{\rm{ID}}}} \in {\mathbb{R} ^{M \times M}}$, respectively. Given an IP address ${q_j}$, let ${H_{{q_j}}}$ and ${D_{{q_j}}}$ be the (\romannumeral1) set of hosts requesting the DNS result of ${q_j}$ and (\romannumeral2) set of domains resolved to ${q_j}$. ${{\bf{S}}_{\rm{IH}}}$ and ${{\bf{S}}_{\rm{ID}}}$ are defined as:
\begin{equation}
    {({{\bf{S}}_{{\rm{IH}}}})_{ij}} := \frac{{|{H_{{q_i}}} \cap {H_{{q_j}}}|}}{{|{H_{{q_i}}} \cup {H_{{q_j}}}|}},{({{\bf{S}}_{{\rm{ID}}}})_{ij}} := \frac{{|{D_{{q_i}}} \cap {D_{{q_j}}}|}}{{|{D_{{q_i}}} \cup {D_{{q_j}}}|}}.
\end{equation}
Furthermore, I combine $\{ {{\bf{S}}_{{\rm{DH}}}},{{\bf{S}}_{{\rm{DI}}}},{{\bf{S}}_{{\rm{IH}}}},{{\bf{S}}_{{\rm{ID}}}}\}$ and ${\bf{P}}$ to construct a similarity-enhanced graph described by the following adjacency matrix:
\begin{equation}
    {\bf{S}} := \left[ {\begin{array}{*{20}{c}} {{{({{\bf{S}}_{{\rm{DH}}}} + {{\bf{S}}_{{\rm{DI}}}})} \mathord{\left/ {\vphantom {{({{\bf{S}}_{{\rm{DH}}}} + {{\bf{S}}_{{\rm{DI}}}})} 2}} \right. \kern-\nulldelimiterspace} 2}}&{{\bf{\hat B}}}\\
    {{{{\bf{\hat B}}}^T}}&{{{({{\bf{S}}_{{\rm{IH}}}} + {{\bf{S}}_{{\rm{ID}}}})} \mathord{\left/ {\vphantom {{({{\bf{S}}_{{\rm{IH}}}} + {{\bf{S}}_{{\rm{ID}}}})} 2}} \right. \kern-\nulldelimiterspace} 2}} \end{array}} \right],
\end{equation}
where all the elements in ${\bf{S}}$ are within the value range $[0, 1]$.

The $K$-order ($K \ge 1$) proximity of a node ${v_i}$ can be used to explore implicit community structures of graphs \cite{qin2018adaptive,qin2019towards,li2019identifying,qin2021dual,qin2023towards,gao2023raftgp}. I apply the matrix-factorization-based objective \cite{qiu2018network} of DeepWalk \cite{perozzi2014deepwalk} to ${\bf{S}}$. Let ${\bf{A}} = {\bf{S}}$ be the adjacency matrix of the heterogeneous graph. This objective aims to factorize the following matrix:
\begin{equation}\label{eq:MF-Obj}
    {\bf{M}} := \log (w{(\frac{1}{K}\sum\nolimits_{k = 1}^K {{{\bf{D}}^{ - 1}}{\bf{S}}} )^k}{{\bf{D}}^{ - 1}}) - \log b,
\end{equation}
where ${\bf{D}} = {\mathop{\rm diag}\nolimits} ({{\deg} (v_1)}, \cdots ,{{\deg} (v_{N + M})})$ is the degree diagonal matrix w.r.t. ${\bf{S}}$; $w = \sum\nolimits_{i = 1}^{N + M} {{{\deg} (v_i)}}$ is defined as the volume of a graph; $K$ and $b$ are pre-set proximity order and number of negative samples.

The low-dimensional embeddings are then derived by fitting the random-walk-based transition probabilities to the sampled random walks equivalent to the following objective \cite{qiu2018network}:
\begin{equation}\label{eq:NE-Obj}
    \mathop {\arg \min }\limits_{{\bf{X}},{\bf{Y}}} {{\mathop{\rm O}\nolimits} _{{\rm{GE}}}}({\bf{X}},{\bf{Y}}) = \left\| {{\bf{M}} - {\bf{X}}{{\bf{Y}}^T}} \right\|_F^2,
\end{equation}
The singular value decomposition (SVD) is then used to get the solution $\{ {{\bf{X}}^*},{{\bf{Y}}^*}\}$ via
\begin{equation}\label{eq:SVD-Solution}
    {{\bf{X}}^*} = {{\bf{U}}_{:,1:d}}\sqrt {{{\bf{\Sigma }}_{1:d}}} ,{{\bf{Y}}^*} = {{\bf{V}}_{:,1:d}}\sqrt {{{\bf{\Sigma }}_{1:d}}},
\end{equation}
with ${\bf{\Sigma }} = {\rm{diag}}({\theta _1},{\theta _2}, \cdots ,{\theta _{N + M}})$ as the diagonal matrix of singular values in descending order. Finally, I adopt ${\bf{X}}^{*}$ as the derived network embeddings, where I treat ${\bf{X}}_{1:N,:}^*$ and ${\bf{X}}_{(N + 1):(N + M),:}^*$ as the domain name embeddings $\{ {{\bf{u}}_1}, \cdots, {{\bf{u}}_N}\}$ and IP address embeddings $\{ {{\bf{v}}_1}, \cdots, {{\bf{v}}_M}\}$, respectively.

In addition, I also incorporate two logistic regression classifiers (i.e., ${C_{{\rm{MDD}}}}$ and ${C_{{\rm{IRE}}}}$) to the unsupervised embedding objective (\ref{eq:NE-Obj}) to support two DNS analysis tasks of (\romannumeral1) malicious domain detection and (\romannumeral2) IP reputation evaluation.

Let $R = \{ {r_{{l_1}}}, \cdots ,{r_{{l_T}}}\}$ and $G = \{ {g_{{l_1}}}, \cdots ,{g_{{l_T}}}\}$ be (\romannumeral1) the result given by classifier and (\romannumeral2) the ground-truth, respectively. The logistic regression classifier can be optimized by minimizing the following binary cross-entropy objective:
\begin{equation}
\resizebox{.90\linewidth}{!}{$
    {{\mathop{\rm O}\nolimits} _{\rm{C}}}(G,R) =  - \sum\nolimits_{i = {l_1}}^{{l_T}} {({g_i}\log (r{}_i) + (1 - {g_i})\log (1 - {r_i}))}
$}.
\end{equation}

To fully utilize the supervised training labels, I further apply the graph regularization technique. A constraint matrix ${\bf{C}} \in {\{ 0, 1\} ^{(N + M) \times (N + M)}}$ is first introduced to represent the training label information.
For a pair of domain names $({p_i},{p_j})$ in the training set, let ${{\bf{C}}_{ij}}={{\bf{C}}_{ji}}=1$ if $({p_i},{p_j})$ are both malicious (or benign) domain names and ${{\bf{C}}_{ij}}={{\bf{C}}_{ji}}=0$ otherwise. Similarly, for a pair of IP addresses $({q_i},{q_j})$ in the training set, let ${{\bf{C}}_{(N+i),(N+j)}}={{\bf{C}}_{(N+j),(N+i)}}=1$ if $({p_i},{p_j})$ have the same reputation level and ${{\bf{C}}_{(N+i),(N+j)}}={{\bf{C}}_{(N+j),(N+i)}}=0$ otherwise. For a given domain name ${p_i}$ and an IP address ${q_j}$ in the training sets, let ${{\bf{C}}_{i,(N + j)}} = {{\bf{C}}_{(N + j),i}} = 1$ if if ${p_i}$ is a benign (or malicious) domain name and ${q_j}$ is with normal (or poor) reputation and ${{\bf{C}}_{i,(N + j)}} = {{\bf{C}}_{(N + j),i}} = 0$ otherwise. Finally, I set ${{\bf{C}}_{i,:}} = {{\bf{C}}_{:,i}} = 0.5$ and ${{\bf{C}}_{(i + N),:}} = {{\bf{C}}_{:,(i + N)}} = 0.5$ for the rest domain names and IP addresses in the validation and test sets. Based on ${\bf{C}}$, the graph regularization objective is then defined as:
\begin{equation}\label{eq:Reg-Obj}
\resizebox{.90\linewidth}{!}{$
    {{\mathop{\rm O}\nolimits} _{\rm{R}}}({\bf{X}},{\bf{C}}) = \frac{1}{2}\sum\nolimits_{i,j} {{{\bf{C}}_{ij}}\left\| {{{\bf{X}}_{i,:}} - {{\bf{X}}_{j,:}}} \right\|_2^2 = {\mathop{\rm tr}\nolimits} ({{\bf{X}}^T}{\bf{LX}})}
$},
\end{equation}
with ${\bf{L}} := {{\bf{D}}_{\mathop{\rm C}\nolimits} } - {\bf{C}}$ as the Laplacian matrix of ${\bf{C}}$ and ${{\bf{D}}_{\rm{C}}} = {\mathop{\rm diag}\nolimits} (\sum\nolimits_j {{{\bf{C}}_{1j}}, \cdots ,\sum\nolimits_j {{{\bf{C}}_{(N + M),j}}} } )$ as the degree diagonal matrix w.r.t. ${\bf{C}}$.

In addition to (\ref{eq:Reg-Obj}), I also develop an end-to-end logistic regression classifier to integrate the supervised training labales. Let $\{ {{\bf{x}}_i}\}  = \{ {{\bf{X}}_{i,:}} = {{\bf{u}}_i}\}  \cup \{ {{\bf{X}}_{N + i,:}} = {{\bf{v}}_i}\}$ to be the set of available DNS entities, including all the domain names and IP addresses. The logistic regression classifier takes each embedding ${{\bf{x}}_{i}}$ as its input and then derives a classification result ${r_i} = \sigma ({{\bf{x}}_i}{\bf{w}} + {\bf{b}})$, with $\{ {\bf{w}}, {\bf{b}}\}$ as learnable parameters. Let $R = \{ {r_1}, \cdots ,{r_{N + M}}\}$ and $G = \{ {g_1}, \cdots ,{g_{N + M}}\}$ be (\romannumeral1) the classification result and (\romannumeral2) the ground-truth w.r.t. DNS entities $\{ {p_1}, \cdots ,{p_N},{q_1}, \cdots ,{q_M}\}$.
I adopt the following binary cross-entropy objective between $G$ and $R$:
\begin{equation}\label{eq:Aux-Clas-Loss}
\resizebox{.88\linewidth}{!}{$
    {{\mathop{\rm O}\nolimits} _{{\rm{AC}}}}(G,R) =  - \sum\nolimits_i {{m_i}({g_i}\log {r_i} + (1 - {g_i})\log (1 - {r_i}))},
$}
\end{equation}
where a mask $m_i$ is integrated to distinguish DNS entities in the training set from those in the validation and test sets. Concretely, ${m_i}=1$ if a DNS entity ${x_i}$ is in the training set and ${m_i}=0$ otherwise.

In summary, one can construct the following semi-supervised objective by combining objectives of the (\romannumeral1) unsupervised graph embedding objective (\ref{eq:MF-Obj}), (\romannumeral2) graph regularization objective (\ref{eq:Reg-Obj}), as well as (\romannumeral3) auxiliary classifier (\ref{eq:Aux-Clas-Loss}):
\begin{equation}\label{eq:obj-semi}
    \mathop {\arg \min }\limits_{{\bf{X}},{\bf{Y}}} {\mathop{\rm O}\nolimits} ({\bf{X}},{\bf{Y}}) = ({{\mathop{\rm O}\nolimits} _{{\rm{GE}}}} + \alpha {{\mathop{\rm O}\nolimits} _{{\rm{AC}}}} + \beta {{\mathop{\rm O}\nolimits} _{\rm{R}}}),
\end{equation}
where $\alpha$ and $\beta$ are hyper-parameters to balance ${\mathop{\rm O}\nolimits}_{\rm{AC}}$ and ${\mathop{\rm O}\nolimits}_{\rm{R}}$.

To obtain the solution of objective (\ref{eq:obj-semi}), I first randomly initialize parameters $\{ {\bf{X}},{\bf{Y}},{\bf{w}},{\bf{b}}\}$ and then apply gradient descent to iteratively update their values until convergence. For the solution of (\ref{eq:obj-semi}) (notated as $\{ {{{\bf{X'}}}^*},{{{\bf{Y'}}}^*}\}$), I use ${\bf{X'}}^{*}$ as the final low-dimensional embeddings for the associated DNS entities. The classification results of malicious domain detection and IP reputation evaluation can then be derived from the end-to-end logistic regression classifier integrated.

\section{Conclusions}\label{sec:con}
In this paper, I proposed a novel JDE method to (\romannumeral1) automatically learn the domain and IP embeddings by jointly exploring the homogeneous and heterogeneous high-order proximities between two types of DNS entities while (\romannumeral2) simultaneously supporting malicious domain detection and IP reputation evaluation. In my future work, I intend to consider the joint optimization of other DNS entities and their associated applications (e.g., device type classification of end hosts). Moreover, to further explore the dynamic natures of DNS query behaviors via existing embedding techniques for dynamic graphs \cite{lei2018adaptive,lei2019gcn,qin2022temporal,qin2023high} is also my next research focus.


\bibliographystyle{IEEEtran}

\end{document}